\documentclass[aps,prl,reprint]{revtex4-1}

\usepackage{graphicx}
\usepackage{amsmath}
\usepackage{amsfonts}
\usepackage{color}

\begin{document}

\title{Anyons and Fractional Quantum Hall Effect in Fractal Dimensions}

\author{Sourav Manna}
\thanks{These authors contributed equally to this work; Current address for Biplab Pal: Department of Physics, Ben-Gurion University, Beer Sheva 84105, Israel.}
\affiliation{Max-Planck-Institut f\"ur Physik komplexer Systeme, D-01187 Dresden, Germany}
\author{Biplab Pal}
\thanks{These authors contributed equally to this work; Current address for Biplab Pal: Department of Physics, Ben-Gurion University, Beer Sheva 84105, Israel.}
\affiliation{Max-Planck-Institut f\"ur Physik komplexer Systeme, D-01187 Dresden, Germany}
\author{Wei Wang}
\thanks{These authors contributed equally to this work; Current address for Biplab Pal: Department of Physics, Ben-Gurion University, Beer Sheva 84105, Israel.}
\affiliation{Max-Planck-Institut f\"ur Physik komplexer Systeme, D-01187 Dresden, Germany}
\author{Anne E. B. Nielsen}
\altaffiliation{On leave from Department of Physics and Astronomy, Aarhus University, DK-8000 Aarhus C, Denmark.}
\affiliation{Max-Planck-Institut f\"ur Physik komplexer Systeme, D-01187 Dresden, Germany}

\begin{abstract}
The fractional quantum Hall effect is a paradigm of topological order and has been studied thoroughly in two dimensions. Here, we construct a new type of fractional quantum Hall system, which has the special property that it lives in fractal dimensions. We provide analytical wave functions and exact few-body parent Hamiltonians, and we show numerically for several different Hausdorff dimensions between $1$ and $2$ that the systems host anyons. We also find examples of fractional quantum Hall physics in fractals with Hausdorff dimension $1$ and $\ln(4)/\ln(5)$. Our results suggest that the local structure of the investigated fractals is more important than the Hausdorff dimension to determine whether the systems are in the desired topological phase. The study paves the way for further investigations of strongly-correlated topological systems in fractal dimensions.
\end{abstract}

\maketitle

Elementary particles that exist in three dimensions are all either bosons or fermions. Nevertheless, Leinaas and Myrheim argued that in two dimensions it is allowed to have particles that are neither bosons, nor fermions, but anyons \cite{leinaas1977theory}. Anyons have unusual exchange properties and can have fractional charge. It was later discovered that anyons are realized physically as quasiparticles in the fractional quantum Hall effect \cite{laughlin1983anomalous,PhysRevLett.53.722}. The investigations of anyons have led to more important insights, including a large development within the description of phase transitions \cite{wen1990topological}, and ideas to use anyons to store and process quantum information in a topologically protected way \cite{RevModPhys.80.1083}.

Among the most important developments in the field are the discoveries of new types of fractional quantum Hall effects and new types of systems where anyons can be realized. This includes the observation of fractional quantum Hall physics in graphene \cite{du2009fractional,bolotin2009observation}, fractional quantum Hall physics in lattice systems \cite{regnault2011fractional}, and generalizations of anyons and the fractional quantum Hall effect to three- or four-dimensional systems \cite{wang2014braiding,4Dexperiment,nandkishore2019fractons}. These developments are important, since each new type of system has its own properties: Graphene gave a relativistic version of the fractional quantum Hall effect, the introduction of lattices eliminated the need for a physical magnetic field, and generalized anyons in three or four dimensions are quite different from anyons in two dimensions.

The possibility of changing the dimension is particularly interesting, since the properties of a system in general depend strongly on the dimension of the system. Introducing fractal structures, it is even possible to consider non-integer dimensions. In the past, classical and single-particle quantum models have been studied on fractal lattices \cite{PhysRevLett.45.855,PhysRevLett.49.1194,PhysRevLett.50.145}, and renewed interest in the topic has appeared in the last few years \cite{PhysRevE.91.012118,van2016quantum,westerhout2018plasmon,pai2018fractalized,
brzezinska2018topology,PhysRevE.98.062114,iliasov2019power,fremling2019chern,pai2019topological}. This interest is, in part, motivated by experimental developments to prepare fractal structures in molecules and nanomaterials \cite{shang2015assembling,kempkes2019design}. Ultracold atoms in optical lattices also provide an interesting direction for realizing fractal models, in particular given the current efforts on creating innovative lattice potentials \cite{PhysRevLett.120.083601}.

The question about topological quantum models on fractals has been taken up recently \cite{pai2018fractalized,brzezinska2018topology, fremling2019chern,pai2019topological}, where noninteracting Chern insulator models have been investigated. The much harder question, whether interacting topological phases, fractional quantum Hall physics, and anyons can exist in fractal space is, however, still unanswered.

In this Letter, we answer this important question in the affirmative by constructing a new type of fractional quantum Hall models, which have the special property that they live in fractal space. We provide analytical wavefunctions and corresponding few-body parent Hamiltonians. We show how to construct anyons in the models and that the anyons are screened and have the correct charge and braiding properties. Our results suggest that anyons and fractional quantum Hall physics can exist in all dimensions between $1$ and $2$. We also find fractional quantum Hall physics on certain fractals with Hausdorff dimension $1$ and below. This first demonstration of topology in interacting systems with fractal dimensions opens up several possibilities for further investigations.

\begin{figure}
\includegraphics[width=\columnwidth,trim=50mm 124mm 60mm 87mm]{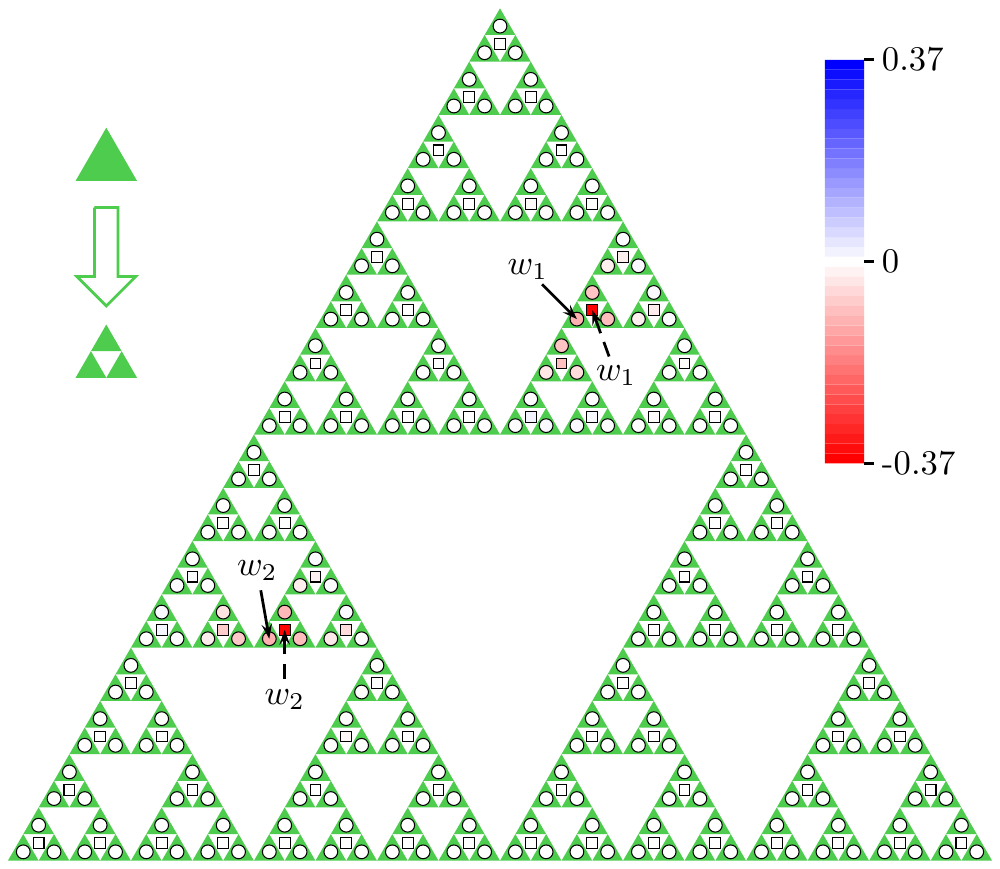}
\caption{The green triangles form a Sierpinski gasket. The operation needed to go from one generation to the next is shown on the left. The considered lattice model has one lattice site on each triangle (generation five (four) is shown with circles (squares), and the solid (dashed) arrows mark $w_k$ for two anyons). The color of the $j$th site shows $\rho_j$, which quantifies how much the anyons affect the particle densities. Here, $q=2$ and $M=30$. It is seen that the anyons are screened and have approximately the same size for generation four and five.}\label{fig:anyon}
\end{figure}

\textit{Fractal lattices.}---Spaces having fractal dimension are realized in fractals, such as the Sierpinski gasket (Fig.\ \ref{fig:anyon}) with dimension $D=\ln(3)/\ln(2)\approx 1.5850$. A Sierpinski gasket of generation 0 is a single triangle, and the Sierpinski gasket of generation $n+1$ is obtained from the Sierpinski gasket of generation $n$ by applying the operation shown in Fig.\ \ref{fig:anyon} to all triangles in the gasket. The full fractal is obtained in the limit of infinite generation.

In any physical system, there is a limit to how small the smallest scales of a fractal can be and a limit to how large the total fractal can be, so the generation of a physical fractal is always finite. As long as we are investigating the fractal at a length scale, which is large compared to the smallest structures of the fractal and small compared to the total size of the fractal, it does, however, not make a difference whether the generation of the fractal is finite or infinite, and the system is effectively in a space with the dimension of the fractal. For the systems considered below, the relevant length scale is set by the typical distance between the particles.

Here, we consider a lattice model on the fractal, where there is one lattice site at the center of each of the smallest triangles. In the limit of large enough generation, it does not make a significant difference, whether we treat each triangle as a triangle or a single point, since the triangles are much smaller than the length scales of interest.

\textit{Quantum states.}---We start our search for anyons in fractal space by constructing fractional quantum Hall states on fractal lattices. The standard fractional quantum Hall effect is realized in a two-dimensional electron gas placed in a strong magnetic field. Just taking a fractional quantum Hall state and restricting the possible particle positions to be on the fractal lattice is not enough to obtain the desired states. The main trick is that we also need to restrict the magnetic flux to only go through the lattice sites. This means that the Gaussian factor, present, e.g., in the Laughlin state, is modified. We can find the appropriate modification by using the conformal field theory approach \cite{moore1991nonabelions,tu2014lattice} to construct the states.

\begin{figure*}
\includegraphics[width=0.32\textwidth]{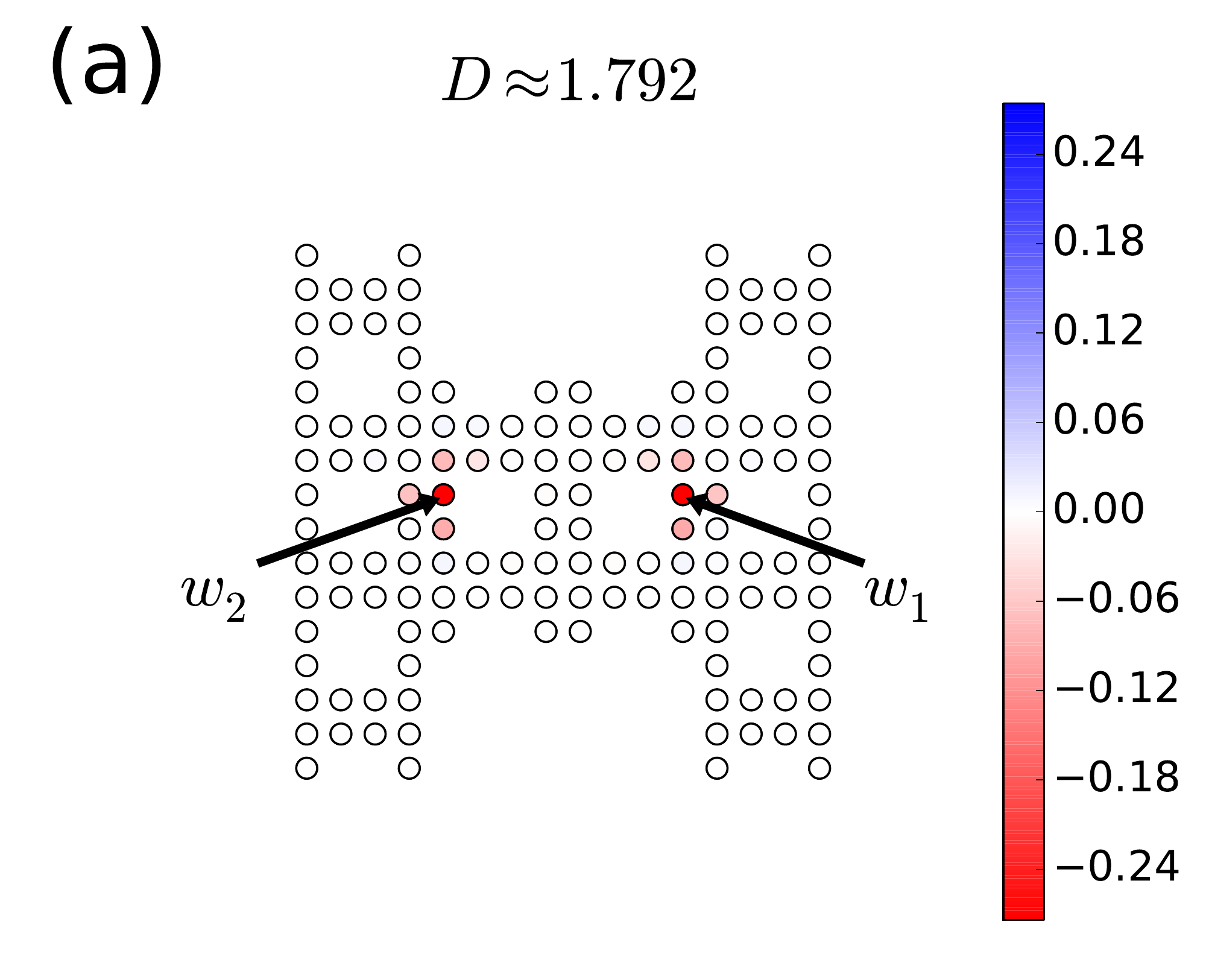}
\includegraphics[width=0.32\textwidth]{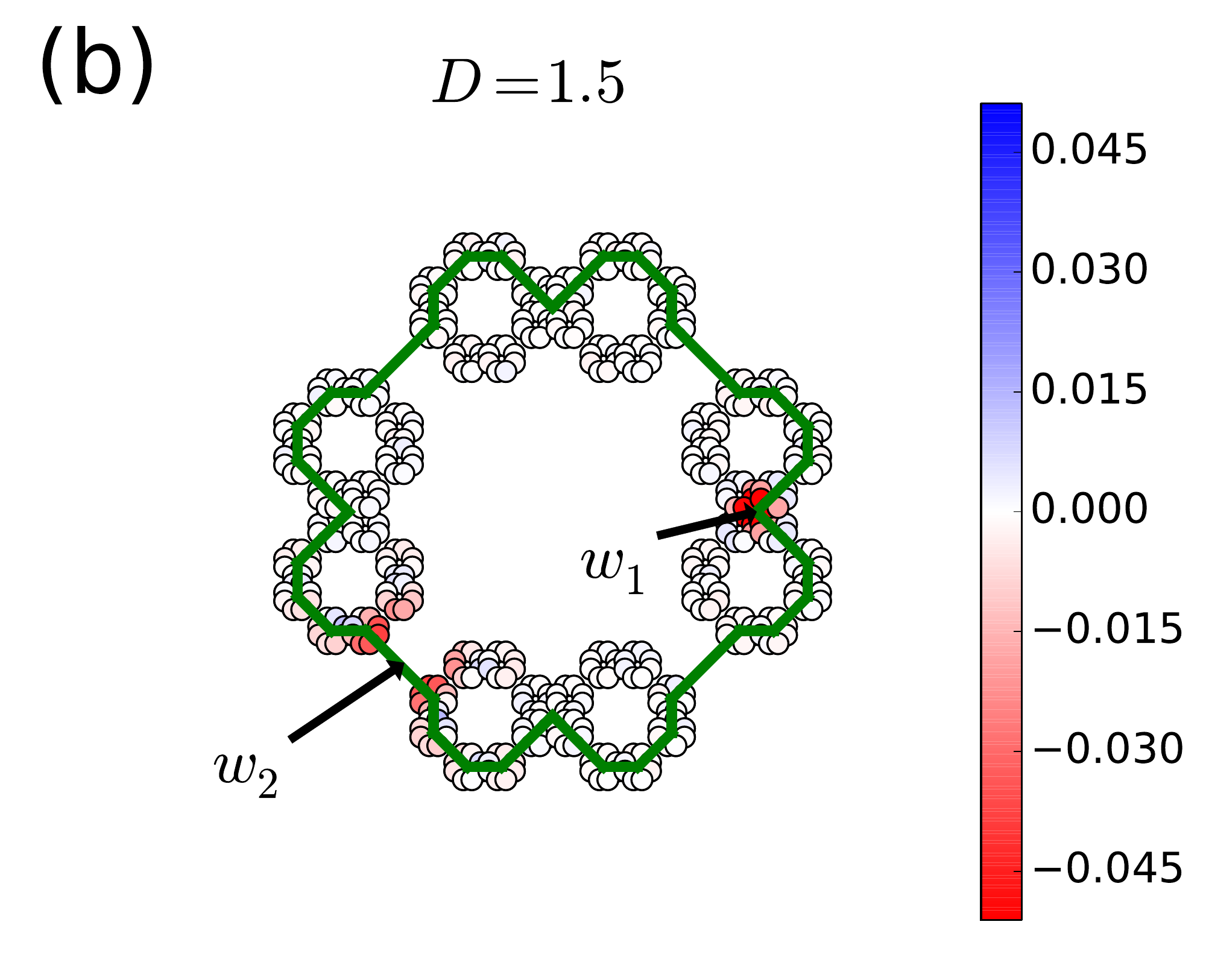}
\includegraphics[width=0.32\textwidth]{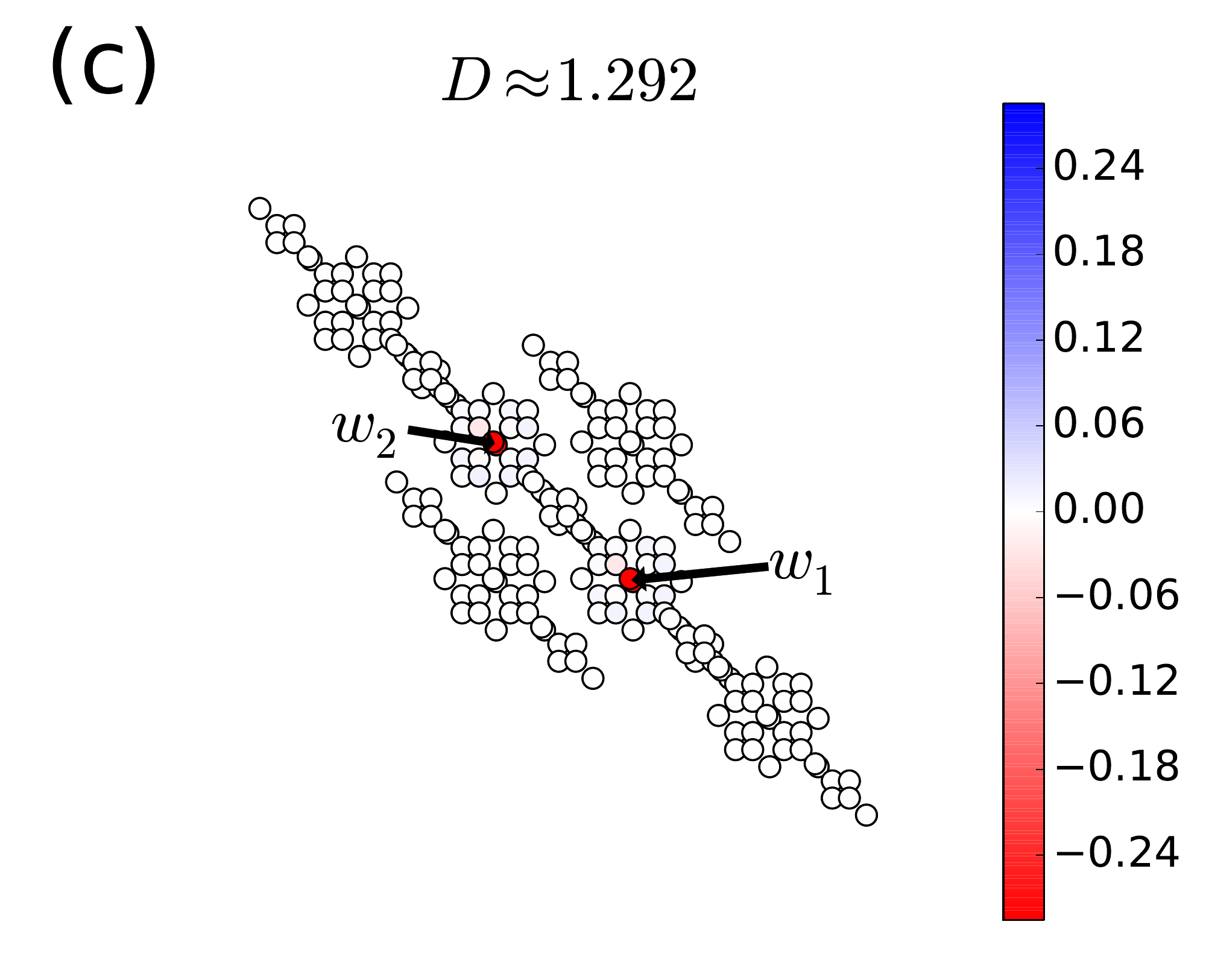}
\caption{Anyons on fractals of different dimensions, generated as shown in Fig.\ \ref{fig:dim}. The color of each lattice site gives $\rho_j$. The anyons are screened in all cases. Here, $q=2$, $M=40$, and the number of sites is $12^2$ in (a), $8^3$ in (b), and $6^3$ in (c). The green line in (b) shows the braiding path chosen in the supplementary videos \cite{sup}.}
\label{fig:screen}
\end{figure*}

We here consider the Laughlin state with $q$ fluxes per particle, where $q$ is a positive integer. A similar construction can be done for Moore-Read states. We associate a vertex operator $V_{n_j}(z_j)={}:e^{i(qn_j-\eta)\phi(z_j)/\sqrt{q}}:$ to each of the lattice sites. Here, $n_j\in\{0,1\}$ is the number of particles on the site, $z_j$ is the position of the site in the plane written as a complex number, $-\eta$ is the magnetic flux through the site, $\phi(z_j)$ is the chiral part of a free massless boson, and $:\ldots:$ means normal ordering.

In the two-dimensional case, we can insert anyons with charge $p_k/q$, where $p_k$ is an integer, into the states by including a vertex operator $H(w_k)={}:e^{ip_k\phi(w_k)/\sqrt{q}}:$ for each of the anyons. Anyons are extended objects, and the $k$th anyon form on the lattice sites surrounding the coordinate $w_k$. We will show below that these operators also produce anyons on the fractal lattice. The Laughlin state on a fractal lattice with $N$ sites, $M$ particles, and $K$ anyons is then defined as $|\psi_{q,K,M}\rangle\propto \sum_{n_1,\ldots,n_N} \langle 0| \prod_{k=1}^K H(w_k) \prod_{j=1}^N V_{n_j}(z_j) |0\rangle |n_1,\ldots,n_N\rangle$, where $|0\rangle$ is the vacuum state. This evaluates to
\begin{multline}\label{wf}
|\psi_{q,K,M}\rangle\propto\sum_{n_1,\ldots,n_N}
\delta_n\prod_{i=1}^N e^{i\phi_jn_j}
\prod_{i=1}^K\prod_{j=1}^N(w_i-z_j)^{p_in_j}\\
\times\prod_{i<j}(z_i-z_j)^{q n_i n_j - n_i\eta - n_j\eta}
|n_1,\ldots,n_N\rangle.
\end{multline}
Here, $\phi_j$ are undetermined phases, which do not influence the results below, and $\delta_n$ is one if $M=\sum_j n_j=(\eta N-\sum_k p_k)/q$ and zero otherwise. Note that the wavefunction has a well-defined limit for $w_k \to z_j$. In all our numerical computations below, we take $p_k=1$. When we go from one generation to the next, we keep the number of particles $M$ and the total flux $-\eta N$ fixed.

\textit{Topological properties.}---We now show that the state \eqref{wf} contains anyons that are screened and have the same charge and braiding properties as anyons in the Laughlin state in two dimensions. The particle density difference
\begin{multline} \rho_j=\langle\psi_{q,K,M-\sum_kp_k/q}|n_j|\psi_{q,K,M-\sum_kp_k/q}\rangle\\ -\langle\psi_{q,0,M}|n_j|\psi_{q,0,M}\rangle
\end{multline}
on site $j$ gives the expectation value of the number of particles on the site, when there are anyons in the system, minus the same quantity, when there are no anyons in the system. The number of particles in the two states is chosen such that the magnetic flux $-\eta$ is the same. The anyons are screened if $\rho_j$ is only different from zero in a small region around each $w_k$. The charge of the $k$th anyon is defined as $Q_k=-\sum_{j\in R_k}\rho_j$, where $R_k$ is a small region around $w_k$, which is large enough to enclose the anyon, but small enough to not enclose other anyons.

We compute $\rho_j$ numerically using the Metropolis Monte Carlo algorithm \cite{metropolis}. Our results for $q=2$ in Fig.\ \ref{fig:anyon} show that the anyons are, indeed, screened and have the expected charge $0.5$. We have chosen the number of particles such that the typical distance between two particles is large compared to the smallest lattice spacing and small compared to the complete fractal. When we increase the generation by one, each lattice site splits into three, but if the generation is already high, there will still be at most one particle on the three sites, since the wavefunction \eqref{wf} is very small if particles are close. The sites therefore effectively act as a single site, and this leads to convergence. It is already visible in the figure that the size of the anyon is practically the same for generation 4 and 5, and the generation is hence large enough to capture the physics of the infinite generation limit.

Computing $\rho_j$ for different positions of the anyons shows that the shape of the anyons depends on the local distribution of lattice sites around the coordinates $w_k$. In all cases, however, the anyons are screened and have the correct charge. It is hence possible to braid them and keep them separated while doing so. We find the same conclusions for $q=3$.

The above computations show that the anyons form on the lattice sites surrounding the coordinates $w_k$. A possible way to move the anyons on the fractal is therefore to vary the parameters $w_k$. A wavefunction $|\psi\rangle$ that depends on a set of parameters $\vec{w}$ acquires the Berry phase $\theta=i\oint_c\langle\psi|\nabla_{\vec{w}}|\psi\rangle d\vec{w}+c.c.$ when moving along a closed path $c$ in the parameter space. Here, we are interested in the statistics $\gamma=\theta_{\textrm{in}}-\theta_{\textrm{out}}$, which is the difference between the Berry phase $\theta_{\textrm{in}}$ when we move the $k$th anyon around the $j$th anyon and the Berry phase $\theta_{\textrm{out}}$ when we move the $k$th anyon along the same path, but with the $j$th anyon outside the path. Inserting \eqref{wf} gives \cite{PhysRevLett.53.722,nielsen2018quasielectrons}
\begin{equation}\label{gamma}
\gamma=i\frac{p_k}{2}\sum_i\oint_c
\frac{\langle n_i\rangle_{\textrm{in}}-\langle n_i\rangle_{\textrm{out}}}{w_k-z_i}dw_k+c.c
\approx\frac{2\pi p_kp_j}{q}.
\end{equation}
The expectation value $\langle n_i\rangle_{\textrm{in}}$ ($\langle n_i\rangle_{\textrm{out}}$) of $n_i$ when the $j$th anyon is inside (outside) the path depends on $w_k$, but if the anyons are screened and well separated throughout, as we have seen above, then the difference $\langle n_i\rangle_{\textrm{in}}-\langle n_i\rangle_{\textrm{out}}$ does not depend on $w_k$ and can be taken outside the integral. Furthermore, the sum over $\langle n_i\rangle_{\textrm{in}}-\langle n_i\rangle_{\textrm{out}}$ over the region inside the path is $-p_j/q$, and the last expression follows. The anyons are hence of Laughlin type.

The supplementary videos \cite{sup} show in detail how $\langle n_i\rangle$ varies with $w_1$ and $w_2$ when braiding two anyons on the fractal in Fig.\ \ref{fig:screen}(b). The first video shows the exchange of two anyons. The Aharonov-Bohm phase is subtracted by subtracting half of the Berry phase when one anyon of twice the charge moves around the same path (second video). By computing integrals similar to the one in \eqref{gamma}, we find the numerical value $\gamma/2=0.50\ \pi$ for the exchange statistics, which agrees with the expected value $\pi/2$.

\begin{figure}
\includegraphics[width=\columnwidth]{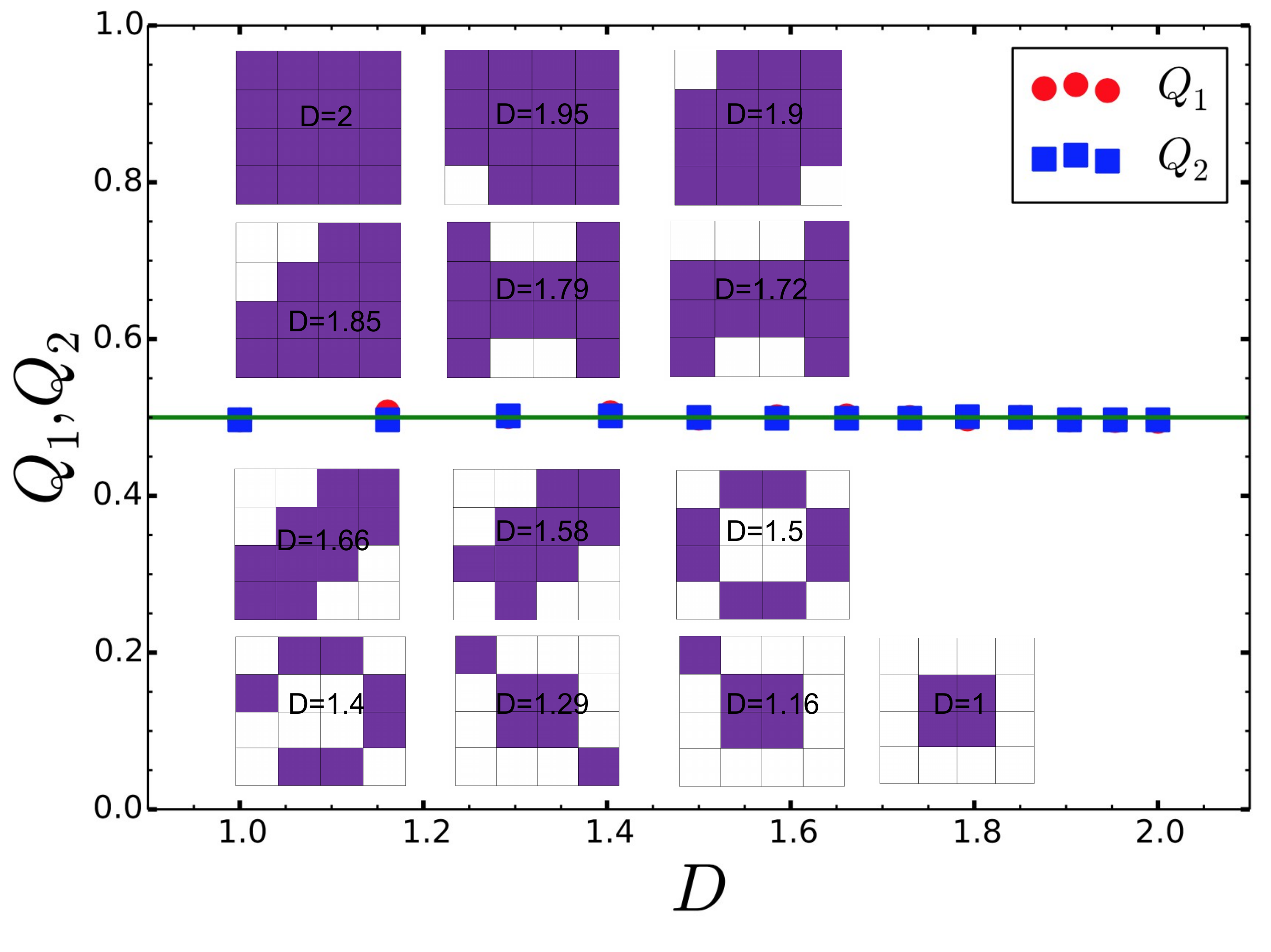}
\caption{We generate fractals of different dimensions by dividing a square into 16 squares, keeping only the squares in purple (insets), and then repeating (the generation is $4$ for $D<1.20$, $3$ for $1.20<D<1.55$, and $2$ for $D>1.55$). In all cases, $q=2$ and $M=40$. The main plot shows the charge of two anyons inserted into the model as a function of the dimension of the fractal (we use the same size of the local region $R_k$ for all cases). The charge is seen to be $0.5$ (marked by the green line) independent of the dimension. This suggest that anyons and the fractional quantum Hall effect can exist in the whole range of dimensions from $2$ to $1$.}\label{fig:dim}
\end{figure}

\textit{Other dimensions.}---To show that the above conclusions hold more generally, we next investigate a family of fractals that allows us to vary the dimension. We start from a square, and we go from one generation to the next by dividing each of the squares present into $L\times L$ squares of equal size and only keeping $U$ of the squares in a particular pattern. The dimension of this fractal is $\ln(U)/\ln(L)$. We study more examples with $L=4$ as shown in Figs.\ \ref{fig:screen} and \ref{fig:dim}. For all the considered cases, we find that the anyons are screened and have a charge of $0.5$. These results suggest that it is, indeed, possible to have anyons and fractional quantum Hall physics in all dimensions between 1 and 2.

\begin{figure}
\includegraphics[width=\columnwidth]{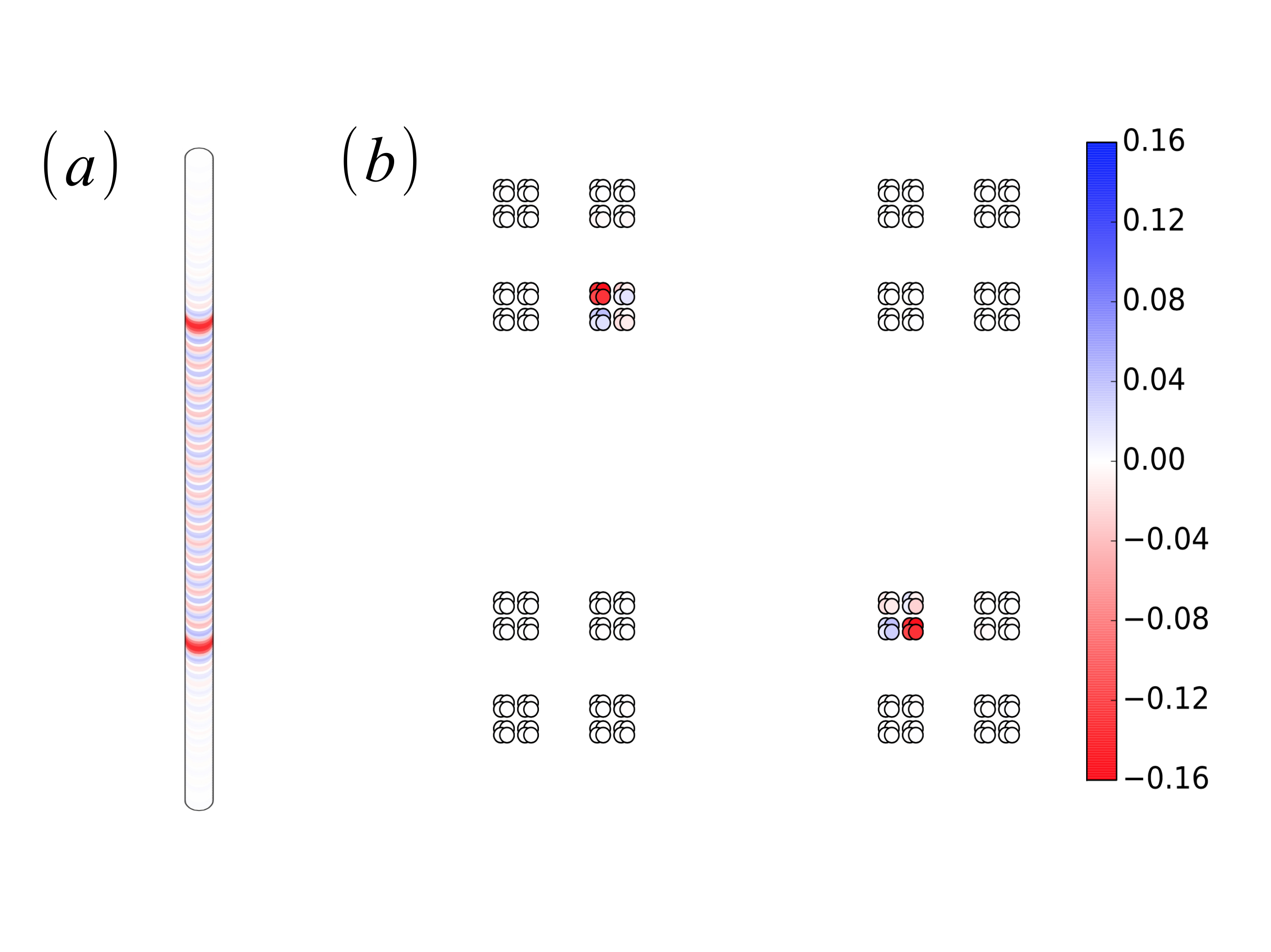}
\caption{(a) If we put our model on a one-dimensional chain, the anyons are not screened. (The circles representing the lattice points overlap each other.) (b) It is, however, possible to have screened anyons and fractional quantum Hall physics in one dimension if we consider the fractal constructed as shown in the lower right inset in Fig.\ \ref{fig:dim}. For both plots, there are $N=4^4$ sites and $M=40$ particles in the system, $q=2$, and the color shows $\rho_j$.}
\label{fig:oneD}
\end{figure}

If we put the model on a one-dimensional line, the anyons are not screened (Fig.\ \ref{fig:oneD}(a)). This is expected, given that this model is critical \cite{tu2014lattice}. Interestingly, however, it is possible to have screened anyons and fractional quantum Hall physics also in one dimension (Fig.\ \ref{fig:oneD}(b)), if we choose the fractal as in Fig.\ \ref{fig:dim}. A similar computation for $L=5$ and $U=4$, with the 4 squares sitting next to each other and forming a square, shows that fractional quantum Hall physics is also possible in dimension $\ln(4)/\ln(5)$. This again shows that the distribution of lattice points in the vicinity of the anyons is more important for the screening than the Hausdorff dimension.

\textit{Exact Hamiltonian.}---So far, we have shown that anyons exist in fractal dimensions, and we have constructed fractional quantum Hall states on fractal lattices hosting anyons. As long as $\eta<1+q/N+\sum_k p_k/N$, it is possible to use the conformal field theory properties of the states to construct a Hamiltonian, which has the state $|\psi_{q,K,M}\rangle$, defined on a general lattice, as ground state \cite{nielsen2018quasielectrons}. This results in the Hamiltonian
\begin{multline}\label{Ham}
H=\sum_{i}\sum_{k(\neq i)}\sum_{j(\neq i)}
\frac{1}{\bar{z}_i-\bar{z}_k} \frac{1}{z_i-z_j} \big[\bar{T}_k^{-1}T_j^{-1} b^\dag_k b_j\\
-\bar{T}_k^{-1}T_i^{-1} b^\dag_k b_i(qn_j-1)
-\bar{T}_i^{-1}T_j^{-1} (qn_k-1)b^\dag_i b_j\\
+|T_i|^{-2}n_i(qn_k-1)(qn_j-1)\big].
\end{multline}
Here, $b_j$ is the operator that annihilates a hardcore boson (fermion) on site $j$, when $q$ is even (odd), $n_j=b_j^\dag b_j$ is the number operator, and
\begin{equation}
T_k=e^{i\phi_k}e^{-i\pi(k-1)}\prod_{i}(w_i-z_k)^{p_i}\prod_j(z_j-z_k)^{1-\eta}.
\end{equation}
We find numerically that the ground state of $H$ is unique, when the number of particles in the system is fixed to $M$. Braiding is done by varying $w_k$, which amounts to varying the strengths of the terms in the Hamiltonian.


\textit{Discussion.}---We have constructed a new type of fractional quantum Hall models that are defined on fractals, and we have shown that anyons can exist in dimensions between $1$ and $2$. We have also shown that fractional quantum Hall physics can appear in systems with Hausdorff dimension $1$ and less than $1$. These results are an important first step that opens up more interesting directions for further studies:

Comparisons between fractional quantum Hall physics in continuous systems and in lattices have revealed important differences and new possibilities \cite{bergholtz2013topological}. The fractal lattices also provide new possibilities, and this strongly motivates a detailed investigation of the interplay between fractional quantum Hall physics and the structure of fractals. It seems particularly promising to study properties that have a strong dependence on the dimension of space, such as transport and entanglement.

The idea presented in this work of restricting the magnetic field to a fractal lattice may give some hints on how to construct fractional Chern insulator type Hamiltonians on fractal lattices. A natural choice would be a model with interactions and complex hopping terms that realize the magnetic field. Such models could pave the way for implementations in ultracold atoms in optical lattices.

Tools to detect topological order in quantum systems often rely on defining the investigated models on closed surfaces. This is, however, not possible for fractals, and the present work motivates the development of additional methods to test for topology.

The present work also motivates the study of strongly-correlated quantum systems in fractal dimensions more generally. The numerical resources needed to study strongly-correlated quantum many-body systems typically grow exponentially with the system size. Two-dimensional systems are therefore difficult to handle. One-dimensional systems are easier, but the physics is often quite different, e.g.\ because there are strong restrictions on the possible directions a particle can move. Fractal dimensions between $1$ and $2$ constitute an intriguing intermediate regime, where interesting physics is likely to happen and little is currently known.

\begin{acknowledgments}
\textit{Acknowledgments.}---The authors thank N. S. Srivatsa for his contributions to computing the plot in Fig.\ \ref{fig:oneD}(a).
\end{acknowledgments}

\end{document}